\documentclass[aps,prb,twocolumn,showpacs]{revtex4}
\usepackage{amssymb}
\usepackage{graphicx}
\usepackage{epsfig}
\usepackage{rotating}
\usepackage{amsmath}

\begin{document}

\title{Stochastic model of self-driven two-species objects in the context of
the pedestrian dynamics.}
\author{Roberto da Silva$^{1}$, Agenor Hentz$^{1}$, Alexandre Alves$^{2}$}
\affiliation{$^{1}$Instituto de Fisica, Universidade Federal do Rio Grande do Sul, Porto
Alegre, RS, Brasil\\
$^{2}$Depto. de Ci\^{e}ncias Exatas e da Terra, Universidade Federal de S%
\~{a}o Paulo, Diadema, SP, Brasil}

\begin{abstract}
In this work we propose a model to describe the statistical fluctuations of
the self-driven objects (species A) walking against an opposite crowd
(species B) in order to simulate the regime characterized by stop-and-go
waves in the context of pedestrian dynamics. By using the concept of
single-biased random walks (SBRW), this setup is modeled both via partial 
differential equations and by Monte-Carlo simulations. The problem is 
non-interacting until the opposite particles visit the same cell of the considered 
particle. In this situation, delays on the residence time of the particles per cell 
depends on the concentration of particles of opposite species. We analyzed 
the fluctuations on the position of particles and our results show a non-regular 
diffusion characterized by long-tailed and asymmetric distributions which is better 
fitted by some chromatograph distributions found in the literature. We also show that
effects of the reverse crowd particles is able to enlarge the dispersion of
target particles in relation to the non-biased case ($\alpha =0$) after
observing a small decrease of this dispersion.
\end{abstract}

\pacs{02.50.-r; 05.40.-a; 89.65.-s}
\maketitle

A large number of stochastic phenomena in literature are related to the
passage of particles through random media generated, for example, by
imperfections of the environment. Examples can be found as disordered linear
chains generated by arbitrary mass and spring-constants \cite{Dyson1953},
random walks in random environments and diffusion (see for example \cite%
{Anshelevich1981,Bernasconi1978,Domingos1999}, charge trapping phenomena
(electron transport) in semiconductor devices (see for example \cite%
{Machlup1954, Kirton1989,
noisesemiconductors,noisesemiconductors2,noisesemiconductors3,noisesemiconductors4}%
), and transport of molecules (chromatography) in Chemistry ( \cite%
{Giddings1950,Giddings1989,Cromatograph}).

On the other hand, the environment can be "perfect" but the particles can
interact by occupying the same region in the space and statistical
fluctuations can be generated not because imperfections but from the
interaction among those particles. The attempts to give explanations about
the dynamics of particles in these situations can be translated into
problems related to the movement of human beings in corridors, crosswalks,
sidewalks and public places in general. But which are the minimal physical
aspects necessary to explain the concentration phenomena of human beings as
a phenomena of concentration of interacting particles or hard bodies?

The corresponding literature in this case is highly concentrated into
evacuation rooms in the context of phenomena related to crowd stampede
induced by panic, driven naturally by the huge importance of the problem.
Results related to optimal strategy for the escape from a smoke-filled room
were, for example, studied in \cite{Delbing2000}. Such problem is deeply
related to occurrences of tragedies. A recent example was the love parade
occurred in the Germany, in 2010 \cite{Krausz2012} when a bad estimate of
number of people in this electronic music party generated more than twenty
deaths and several injured. However no modeling were developed to study, for
example, statistical effects of people in contrary flux although some works
have already explored the problem under other point of view. In \cite%
{Zhang2014} comparisons of intersecting pedestrian flows were analyzed based
on experiments.

For the modeling of pedestrian dynamics some models were explored in order
to study the influence of several effects for the interaction of pedestrian
on the resulting velocity-density relation. An interesting point on those
models is the transition from laminar flow regime to the regime of known
stop-and-go wave phenomena that occurs when density of pedestrian increases
above critical value of density (see for example \cite{Helbing2007} and \cite%
{Portz2010})

We believe that peculiar characteristics of these \textquotedblleft
crowd\textquotedblright\ effects can be modeled in terms of a simple
stochastic approach by adding some important ingredients. Let us imagine a
problem that considers a straight line divided into cells where there are
two crowds, denoted by groups (or species) $A$ and $B$. Without loss of
generality all participants of group $A$ are initially placed in the far
left-hand cell and all participants of group $B$ are initially placed in the
far right-hand cell.

The idea is that group $A$ has the aim to arrive at the far right-side
(starting point of species $B$, target of specie $A$) and the group $B$ to
arrive at the far left-side (starting point of species $A$, target of specie 
$B$). Since the population $A $ and $B$ disturb each other, a natural
approach for the problem is to describe it in terms of modified random walks
in which there are only two possibilities of movement to each element at any
given time-step: to stay still or take a step to the nearest neighbor cell
towards its target.

Naturally, many similar phenomena can be imagined even from a theoretical
point of view: the flux of molecules in chromatographic columns or the
electronic transport in non-homogeneous media. Surely we are not simply
comparing human beings crossing a corridor with molecules crossing a
chromatographic column, but we are calling the attention to some similar
aspects from concentration phenomena that can also ocurr when many bodies
directed by some field (for example to arrive at bus or subway) cross
reverse crowds.

In the context of mean field regime, Montroll and West \cite{MontrollandWest}
described a problem which has some relation to the present one: the problem
of \textquotedblleft clannish random walks\textquotedblright. In that case
two species, $A$ and $B$, execute concurrent random walks characterized by
the intensification of the clannishness of the members of one species as a
function of the concentration of the other species. However our problem has
the opposite sense of that one arising from clannishness since in our model
a particle can still in same cell or step towards for the next cell
(directed random walk) in the direction of its target and its aim is simply
to cross de corridor and not to make part of a clan.

By capturing this idea, we propose in this paper a model where species $A$($%
B $), with opposite targets, has a decrease in the probability to step to
the right(left) proportional to the concentration of other particles $B$($A$%
) offering resistance during the passage. Therefore we define the following
probabilities:

\begin{equation}
\begin{array}{lll}
\Pr^{(A,B)}(k\rightarrow k\pm 1|\rho _{B,A}) & = & p-\alpha (1-\rho _{A,B})
\\ 
& = & p-\alpha (1-\rho )-\alpha \rho _{B,A} \\ 
& = & 1-\Pr^{(A,B)}(k\rightarrow k|\rho _{B,A})%
\end{array}
\label{prob}
\end{equation}

Here \ $\Pr^{(A,B)}(k\rightarrow k\pm 1|\rho _{B,A})$ denotes the
probability of particle $A[B]$ at position $\ x=ka$ to move to the nearest
neighbor $(k+1)a\ [(k-1)a]$ cell given that the concentration of species $A$
($B$) in its cell is $\rho _{A}$ ($\rho _{B}$) where $\rho =\rho
_{A}+\rho_{B}$. Naturally, $\Pr^{(A,B)}(k\rightarrow k|\rho _{B,A})$ denotes
the probability of this particle to stay still at the cell under same
restrictions. Here $p=1-q$ define the probability of a given particle to
move to the nearest neighbor cell towards its respective target when the
current cell contains only particles of its specie. The parameter $\alpha $,
such that $0<\alpha <p<1$ measures the resistence level bias which, as well
as $p$, at least in a first analysis, assumes the same value for all
players. The problem is symmetric, and therefore we use the referential of
the particles $A$, since particles $B$ present a similar behavior.

We define $n(ka,l\tau )=n_{k,l}$ as the number of particles at cell $k$
after $l$ steps. Here $a$ is the size of each cell and $\tau $ the
time-step. In mean field regime we can write the recurrence relation $%
n_{k,l+1}=\Pr^{(A)}(k-1\rightarrow k|\frac{n(k-1,l)}{N})\cdot
n_{k-1,l}+\Pr^{(A)}(k\rightarrow k|\frac{n(k,l)}{N})\cdot n_{k,l}$ and so

\begin{equation}
\begin{array}{c}
n_{k,l+1}=\left[ p-\alpha \left( 1-\frac{n_{k-1,l}}{N}\right) \right]
n_{k-1,l}+ \\ 
\left[ q+\alpha \left( 1-\frac{n_{k,l}}{N}\right) \right] n_{k,l}%
\end{array}
\label{recurrence_relation_1}
\end{equation}

So $n_{k,l+1}=\left[ p-\alpha \left( 1-\frac{n_{k-1,l}}{N}\right) \right]
n_{k-1,l}+\left[ 1-p+\alpha \left( 1-\frac{n_{k,l}}{N}\right) \right]
n_{k,l} $, and so 
\begin{equation}
\begin{array}{c}
n_{k,l+1}-n_{k,l}=-p(n_{k,l}-n_{k-1,l})+ \\ 
\alpha \left[ \left( 1-\frac{n_{k,l}}{N}\right) n_{k,l}-\left( 1-\frac{
n_{k-1,l}}{N}\right) n_{k-1,l}\right]%
\end{array}
\label{recurrence_relation_2}
\end{equation}

Here the number $N$ deserves some discussion. In the strict sense of random
walkers, and not for the adaptation of this model to explain the phenomena
of human beings crossing corridors, we postulate that the number of walkers
in each cell in mean-field approximation is supposed to remain constant,
hence $N=\left\langle n\right\rangle =\rho a$, with $\rho =\rho _{A}+$ $\rho
_{b}$ being the total density of particles per cell. However, for computer
simulations we can study the problem by using the $\rho _{A}$ and $\rho _{b}$
correctly calculated per cell and compare it with the results obtained for
the mean field approximation.

\begin{figure}[h]
\begin{center}
\includegraphics[width=0.75\columnwidth]{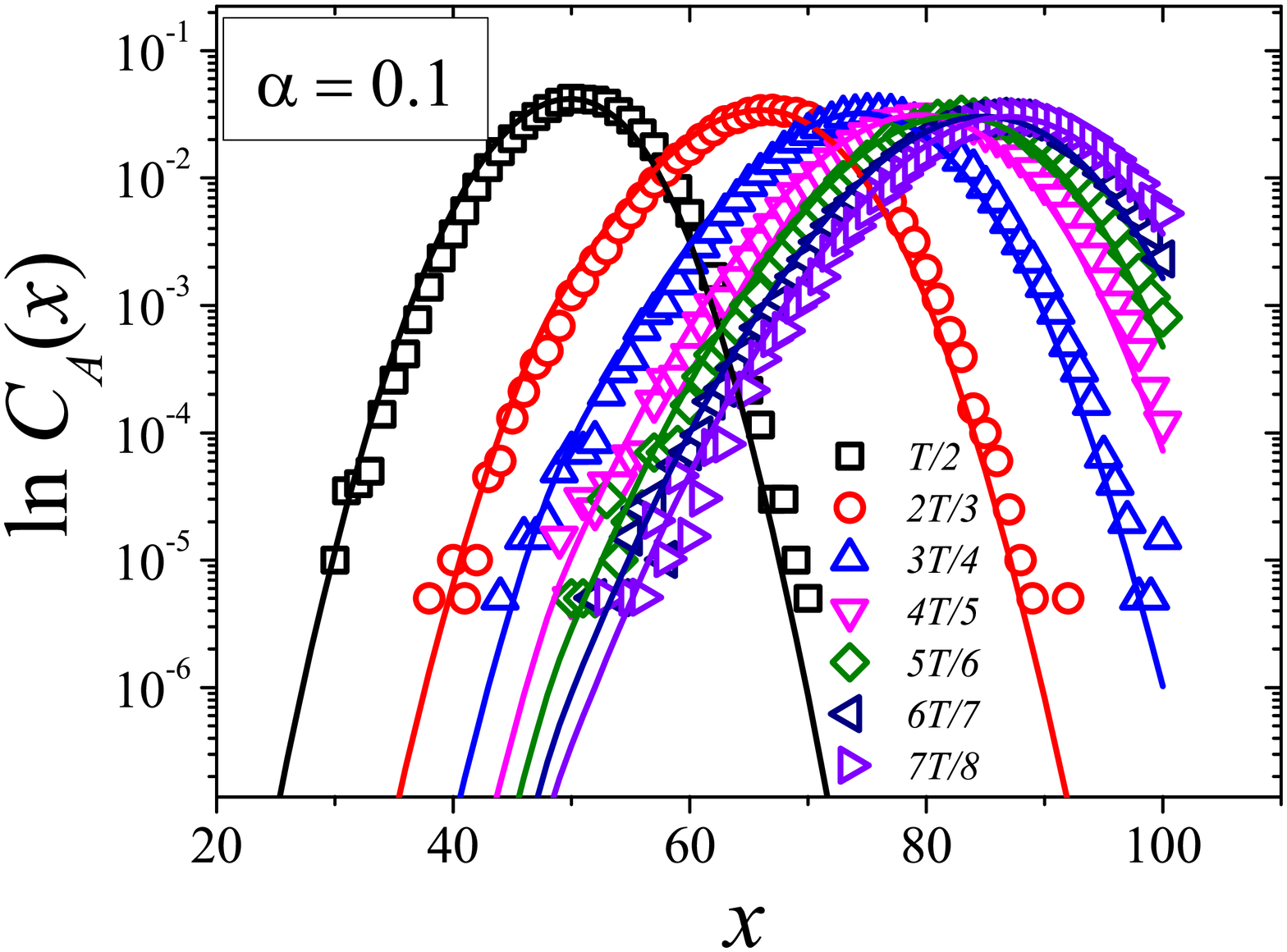} %
\includegraphics[width=0.75\columnwidth]{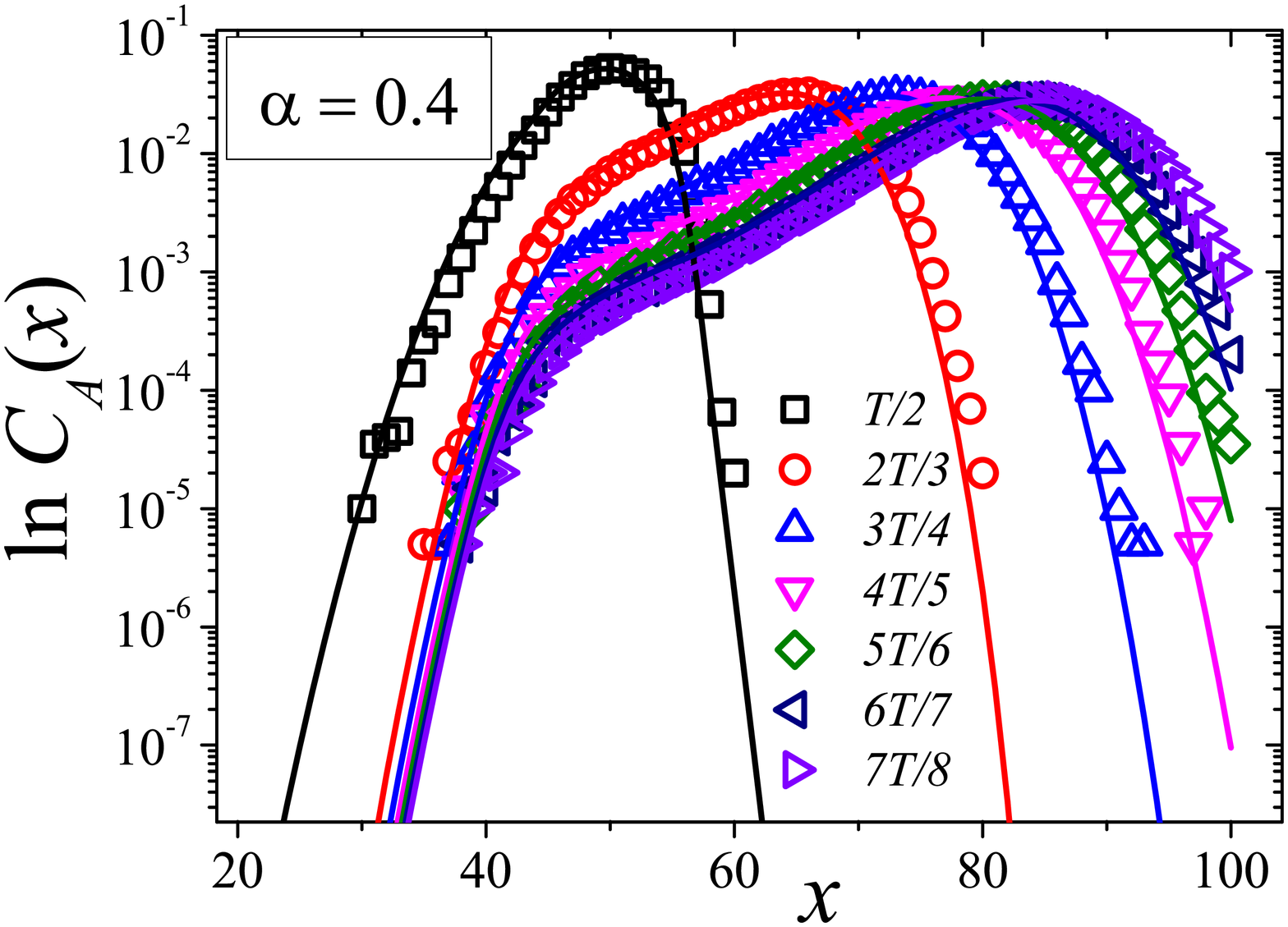}
\end{center}
\caption{Distribution of particles $A$ for $p=1/2$. We can observe a
stronger deformation from the quadratic behavior in mono-log scale (Gaussian
behavior) for $\protect\alpha =0.4$ than for $\protect\alpha =0.1$.
Continuous curves correspond to the solution of PDE while points correspond
to MC simulations.}
\label{Fig:concentration}
\end{figure}

So by completing our mean-field results, since $n_{k,l+1}-n_{k,l}$ in first
approximation is $N\tau \frac{\partial c_{A}}{\partial t}$, $%
(n_{k,l}-n_{k-1,l})$ is $Na\frac{\partial c_{A}}{\partial x}$ and $\left( 1- 
\frac{n_{k,l}}{N}\right) n_{k,l}-\left( 1-\frac{n_{k-1,l}}{N}\right)
n_{k-1,l}$ is $Na\frac{\partial \left[ c_{A}(1-c_{A})\right] }{\partial x}$
we have $\tau \frac{\partial c_{A}}{\partial t}=-ap\frac{\partial c_{A}}{
\partial x}+\alpha a\frac{\partial \left[ c_{A}(1-c_{A})\right] }{\partial x}
$ which results in%
\begin{equation}
\frac{\partial c_{A}(t,x)}{\partial t}=-A_{1}\frac{\partial c_{A}(t,x)}{
\partial x}-A_{2}c_{A}(t,x)\frac{\partial c_{A}(t,x)}{\partial x}
\label{independent}
\end{equation}%
where $A_{1}=\lim_{a,\tau \rightarrow 0}\frac{a}{\tau }(p-\alpha )\ $ and $%
A_{2}=\lim_{a,\tau \rightarrow 0}2\alpha a/\tau$.

An "ansatz" for the solution is $c_{A}(t,x)=f(x-t(A_{1}+A_{2}c_{A}))$.
Deriving both sides with respect to $t$ one obtains $\frac{\partial c_{A}}{%
\partial t}=-(A_{1}+A_{2}c_{A}+A_{2}t\frac{\partial c_{A}}{\partial t}%
)f^{\prime }(z)$ \ where $z=$ $x+t(A_{1}+A_{2}c_{A})$. Similarly, deriving
both sides with respect to $x$ one obtains

\begin{equation}
\frac{\partial c_{A}}{\partial x}=(1-A_{2}t\frac{\partial c_{A}}{\partial x}%
)f^{\prime }(z).  \label{x}
\end{equation}

From Eq. \ref{x} and the expression for $\partial c_{A}/\partial t$ we have $%
(1-A_{2}t\frac{\partial c_{A}}{\partial x})\frac{\partial c_{A}}{\partial t}%
=-(A_{1}+A_{2}c_{A}+A_{2}t\frac{\partial c_{A}}{\partial t})\frac{\partial
c_{A}}{\partial x}$, which leads to Eq. \ref{independent}. So given an
initial distribution $c_{A}(0,x)=f(x)$ the time evolution of concentrations
can be determined. However for the present case $f(x)=\delta (x)=\frac{1}{%
2\pi }\int_{-\infty }^{\infty }e^{ikx}dk$ such solution is not suitable%
\footnote{%
We are better exploring the possible interesting analytical solutions in
other contribution in progress. In this same contribution we perform a
classification of the problem and its ramifications.}. Moreover, by keeping $%
N$ constant, as suggested by Montroll in the context of clannish random
walks, is not interesting since given our initial condition such solution
does not capture the essential behavior of the problem. So changing $N$ by
its real value: $n_{k,l}+m_{k,l}$ where $m_{k,l}$ denotes the number of
particles of specie $B$ in the $k$-th cell at time $t=l\tau $, instead of
Eq. \ref{recurrence_relation_2} we have now the concentrations of $A$ and $B$
described by two coupled equations: 
\begin{equation}
\begin{array}{c}
m_{k-1,l+1}=m_{k-1,l}+p(m_{k,l}-m_{k-1,l})- \\ 
\alpha \left[ \left( 1-\frac{m_{k,l}}{n_{k,l}+m_{k,l}}\right) m_{k,l}-\left(
1-\frac{m_{k-1,l}}{n_{k-1,l}+m_{k-1,l}}\right) m_{k-1,l}\right] ,%
\end{array}
\label{recurrence_relation_3}
\end{equation}%
and 
\begin{equation}
\begin{array}{c}
n_{k,l+1}=n_{k,l}-p(n_{k,l}-n_{k-1,l})+ \\ 
\alpha \left[ \left( 1-\frac{n_{k,l}}{n_{k,l}+m_{k,l}}\right) n_{k,l}-\left(
1-\frac{n_{k-1,l}}{n_{k-1,l}+m_{k-1,l}}\right) n_{k-1,l}\right] .%
\end{array}
\label{recurrence_relation_4}
\end{equation}

We end up with two coupled PDE equations which describe the problem instead
of the uncoupled equation \ref{independent}:

\begin{eqnarray}
\frac{\partial c_{A}}{\partial t} &=&-k_{1}\frac{\partial c_{A}}{\partial x}%
+k_{2}\frac{\partial }{\partial x}\left( \frac{c_{A}c_{B}}{c_{A}+c_{B}}%
\right)  \label{coupled} \\
&&  \notag \\
\frac{\partial c_{B}}{\partial t} &=&k_{1}\frac{\partial c_{B}}{\partial x}%
-k_{2}\frac{\partial }{\partial x}\left( \frac{c_{A}c_{B}}{c_{A}+c_{B}}%
\right)  \notag
\end{eqnarray}%
resulting in $\frac{\partial (c_{A}+c_{B})}{\partial t}=-k_{1}\frac{\partial
(c_{A}-c_{B})}{\partial x}$, where $k_{1}=\lim_{a,\tau \rightarrow 0}\frac{a%
}{\tau }p$ and $k_{2}=\lim_{a,\tau \rightarrow 0}\frac{a}{\tau }\alpha $.

We numerically solved the recurrences \ref{recurrence_relation_3} and \ref%
{recurrence_relation_4} and concurrently we also performed Monte Carlo
simulations of the problem monitoring 3 aspects of the $c_{A}$ distribution
(since $c_{B}$ presents symmetric behavior):

\begin{enumerate}
\item Since $c_{A}$ reflects the residence time of particles in the medium,
which are the aspects of distribution $c_{A}$?

\item Which are the temporal aspects $\left\langle x\right\rangle $ and $%
\sigma(t)=\left( \left\langle x^{2}(t)\right\rangle -\left\langle
x(t)\right\rangle ^{2}\right) ^{1/2}$ as well as of the skewness and
kurtosis?

\item What about the effects of $\alpha $?
\end{enumerate}

Our MC simulations were performed synchronously, in the context of cellular
automata, i.e., all particle actions are taken simultaneously. The
simulations have the initial condition as required by our aims: a given
number of particles $A$ ($N_{part}$) in the position $x=0$ and the same
number of particles $B$ in the position $x=aN_{cel}$, where $N_{cel}$ is the
number the total number of cells. The other cells are initially empty which
means an initial delta distribution as desired. On the other hand the
similar condition for the integration of equations \ref%
{recurrence_relation_3} and \ref{recurrence_relation_4} is to set: $%
n_{0,0}=1 $ and $m_{N_{cel},0}=1$ and $n_{k,0}=m_{k,0}$ for all $k\neq 0$
and $k\neq N_{cel}$.

Here, we keep our analysis restricted to the fundamental case $p=1/2$. It is
important to emphasize that the analysis for $p\neq 1/2$ follows in a
similar fashion respecting the natural scale of the problem: $T=N_{cel}/p$.
Fig.\textbf{\ }\ref{Fig:concentration}\textbf{\ }shows the distribution of
particles for two different values of the resistance parameter $\alpha =0.1$
(upper panel) and $\alpha =0.4$ (lower panel). We analyzed the concentration
of particles at 8 different times ($t=(k-1)T/k$, for $k=1,2,..,8$. The
reason is simple: since the particle $A$ leaps to the right with probability 
$p$, the probability that $m$ trials are necessary for a given particle to
jump $N_{cels}$ is a negative binomial distribution: $p_{m,N_{cel}}=\binom{%
m+N_{cel}-1}{N_{cel}-1}p^{N_{cel}}(1-p)^{m-N_{cel}}$ which leads to $%
T=\left\langle m\right\rangle =\sum_{m=N_{cel}}^{\infty }m\
p_{m,N_{cel}}=N_{cel}/p$. This is the average crossing time for a particle
when $\alpha =0$, which is always larger than $T_{\alpha \neq 0}$ and
therefore our baseline. We used $N_{cel}=100$ for all experiments in this
paper by default, except when stated otherwise.

For a small value ($\alpha =0.1$) we observe something next of a Gaussian
behavior (quadratic function in mono-log scale). For ($\alpha =0.4$) we
observe a strong deformation of such behavior. The continuous curves
correspond to the solution of PDE while the points correspond to MC
simulations.

\begin{figure}[h]
\begin{center}
\includegraphics[width=0.75\columnwidth]{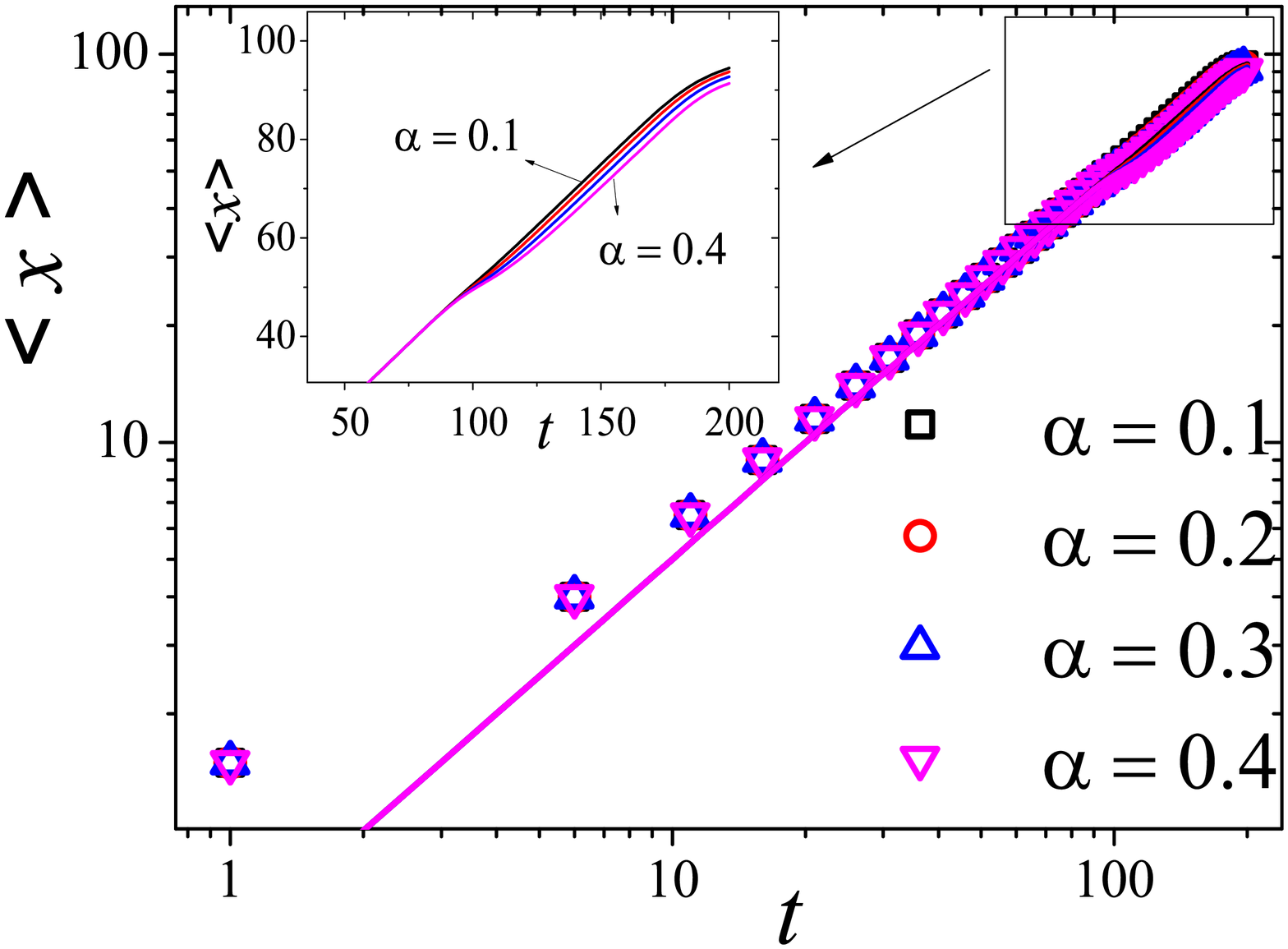} \includegraphics[width=0.75%
\columnwidth]{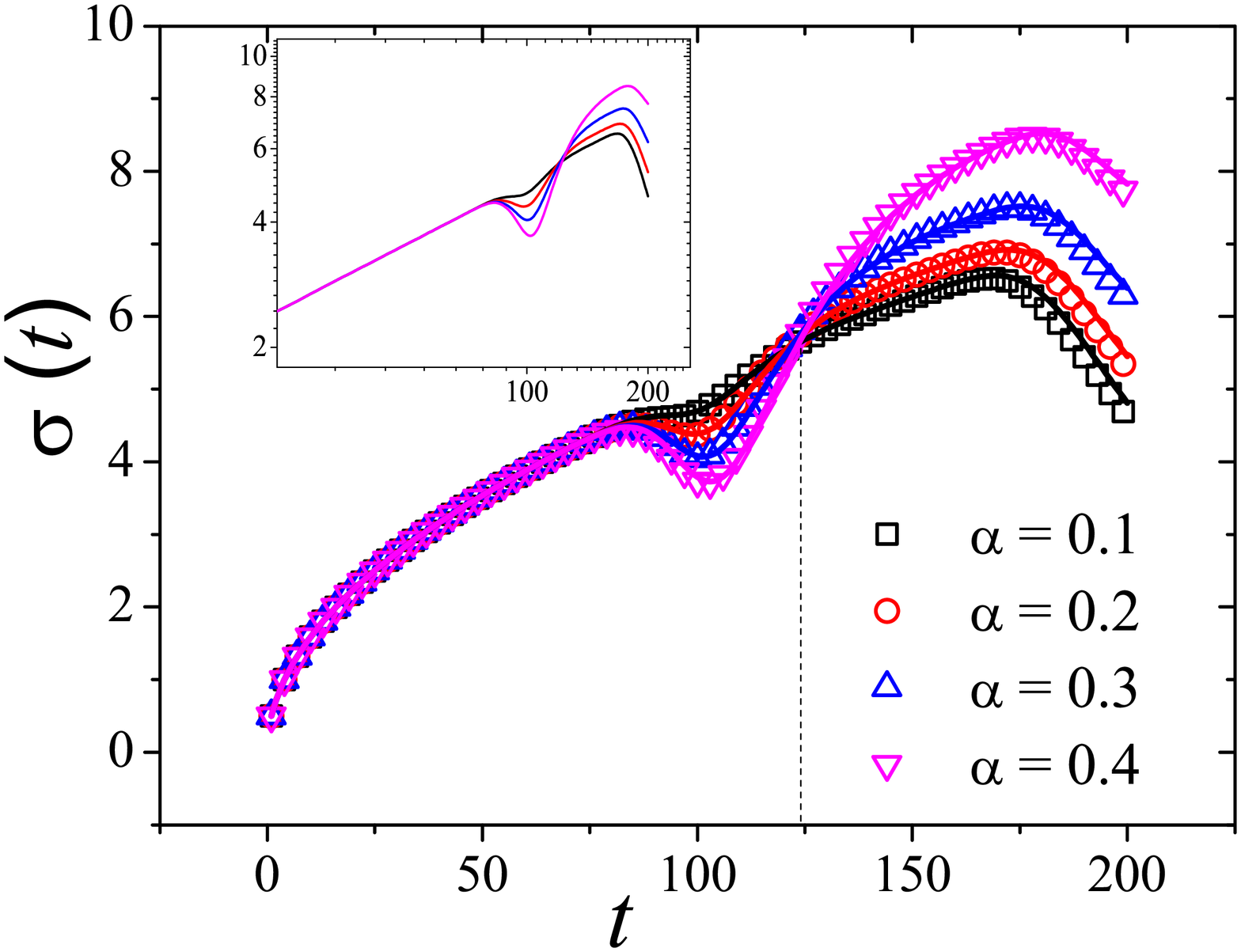}
\end{center}
\caption{Temporal description of average (upper panel) and variance (lower
panel) of particle position along the environment. The inset plot in upper panel 
describes  a zoom of selected region and in the lower panel one corresponds to 
the log-log plot of the dispersion vs time.}
\label{average_variance}
\end{figure}

In order to describe with more details the dynamic behavior of the particle
distribution along the environment, we compute the average position, its
standard deviation as well as two other important quantities: the skewness
and kurtosis of particle distribution for $\alpha =0.1$, $0.2$, $0.3 $, and $%
0.4$, which can be calculated via the equations: $\left\langle
x(t)\right\rangle =\sum_{x=0}^{N_{cel}}x\
c_{A}(x,t)/\sum_{x=0}^{N_{cel}}c_{A}(x,t)$,$\ \sigma (t)=$ $\sqrt{%
\left\langle x^{2}\right\rangle -\left\langle x\right\rangle ^{2}}=\left[
(\sum_{x=0}^{N_{cel}}x^{2}\
c_{A}(x,t)/\sum_{x=0}^{N_{cel}}c_{A}(x,t))^{2}-\left\langle x\right\rangle
^{2}\right] ^{1/2}$, $skew(t)=\frac{1}{\sigma (t)^{3}}%
\sum_{x=0}^{N_{cel}}(x-\left\langle x\right\rangle )^{3}\ c_{A}(x,t)$ and
excess kurtosis $kurt(t)=\frac{1}{\sigma (t)^{4}}\sum_{x=0}^{N_{cel}}(x-%
\left\langle x\right\rangle )^{4}\ c_{A}(x,t)-3$. The temporal monitoring of
such quantities has showed important aspects about the fluctuations of the
particle concentration.

For example in Fig. \ref{average_variance}, the plots (a) and (b) show
respectively the behavior of the average ($\ \left\langle x(t)\right\rangle
\ $) and variance $\sigma (t)$ of the particle position as a function of $t$%
. The inset plots show respectively the same plots in log-log scale. We can
observe the particle position (plot a) has the same behavior for all values
of $\alpha $ up to a branching point, which corresponds to the moment where
particles $A$ and $B$ start encountering each other on their way to their
targets. From that, the bigger is $\alpha $, the bigger is the decrease of $%
\left\langle x(t)\right\rangle $ in relation to the non-interacting problem (%
$\alpha =0$). More interesting effects emerge from the behavior of the
variance (plot b). In this case particles $A$ have a decreasing in
dispersion as they start to interact with particles $B$, then increasing
again arriving at a "iso-variance" point, i.e., the variance of the particle
distribution is the same independent of the $\alpha $ value. Following, the
dispersion becomes higher than the non-interacting case arriving at a peak
that is higher as $\alpha $ increases. After this point the particles start
to leave the environment and dispersion starts to decrease again.

\begin{figure}[h]
\begin{center}
\includegraphics[width=0.75\columnwidth]{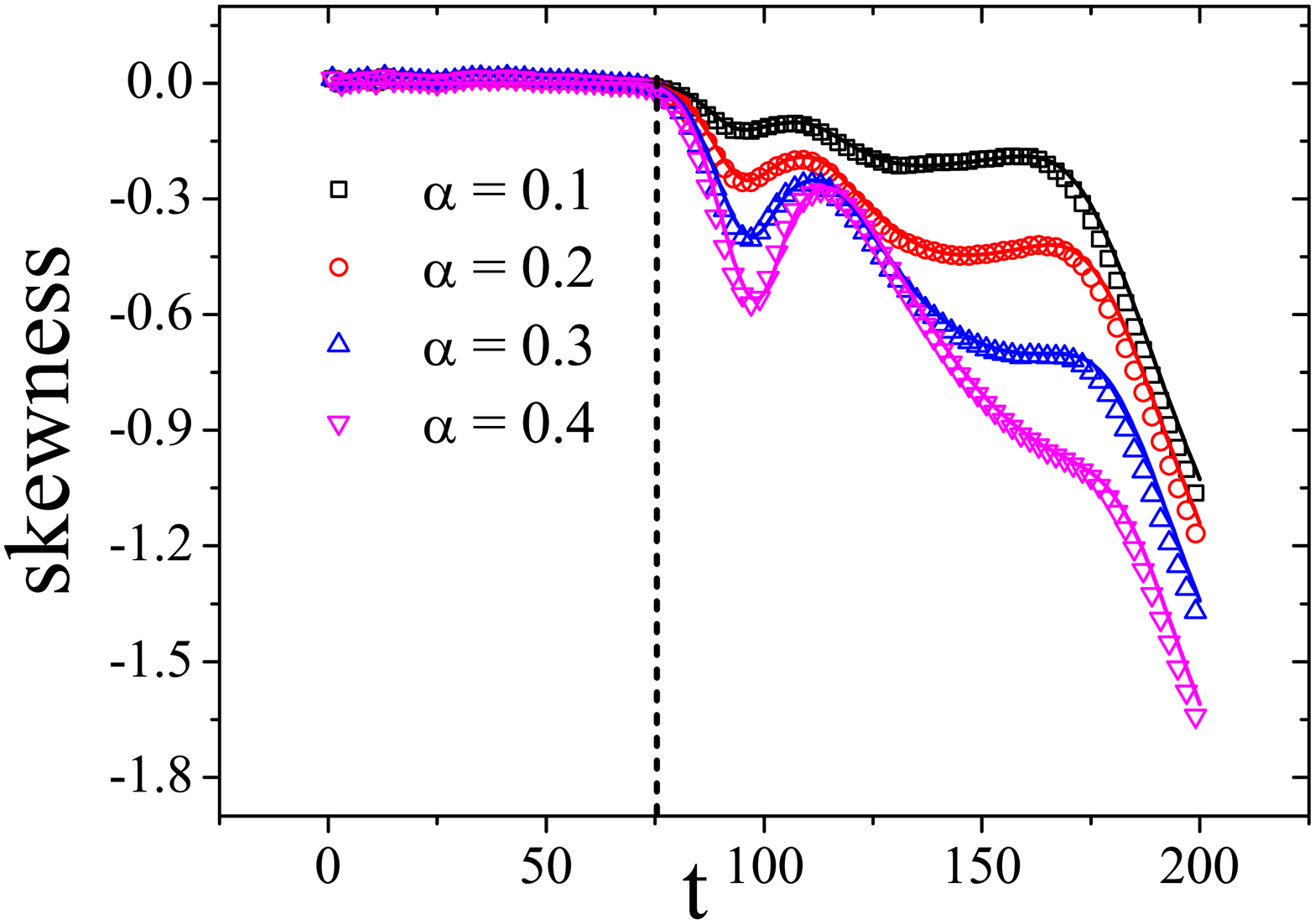} \includegraphics[width=0.75%
\columnwidth]{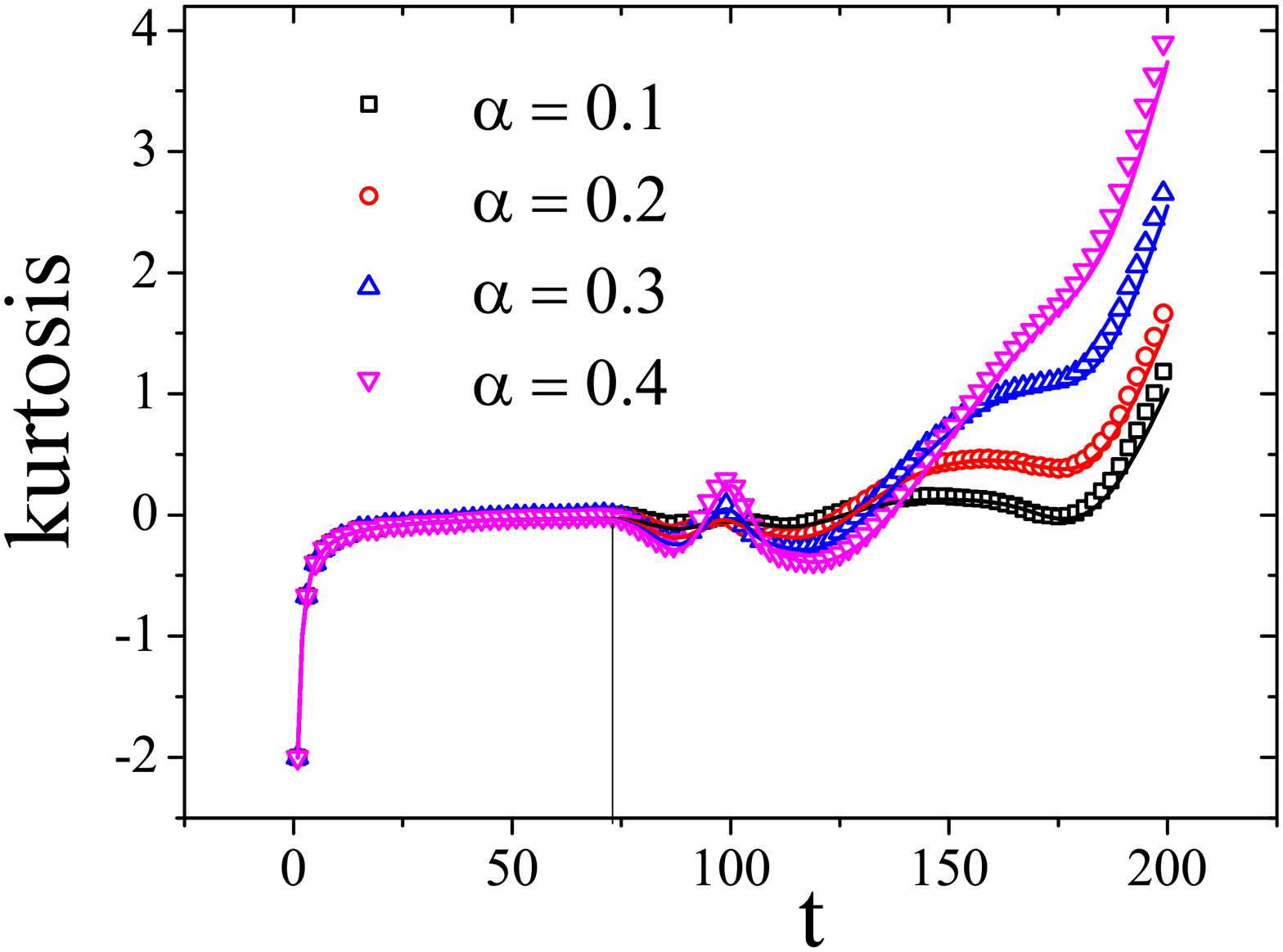}
\end{center}
\caption{Temporal description of skewness (upper panel) and kurtosis excess
(lower panel) of particle position along the environment. }
\label{skew_kurtosis}
\end{figure}

Following we measured the asymmetry and tail weight of the particle
distribution. The results are surprising and very interesting (see Fig. \ref%
{skew_kurtosis}). In the upper panel of Fig. \ref{skew_kurtosis} we firstly
check the peculiar behavior of skewness. After the confrontation of
particles we have a cross-over from $skew=0$ (Gaussian) to a non-symmetrical
distribution $skew<0$. A general pattern is observed for all values of $%
\alpha $: initial valley followed by a shoulder and a subsequent decrease
(this last one due to the scape of particles from the confined environment).

In the lower panel we observe an analogous cross-over for the behavior of
the kurtosis ($kurt=0$ to positive kurtosis). The kurtosis has a little peak
around $t=100$ MCsteps, coincidently the same point at which the skewness
presents the valley. After we check a increase of the tail, which is longer
as $\alpha $ increases. By performing the simulations for different values
of system with the same $p$ value, we observe the maximum of the kurtosis
peak and the minimum of the skewness valley occur at same time of evolution
according to the scaling $\tau =N_{cel}$, which is intuitively expected.
\qquad \qquad

\begin{figure}[h]
\begin{center}
\includegraphics[width=0.75\columnwidth]{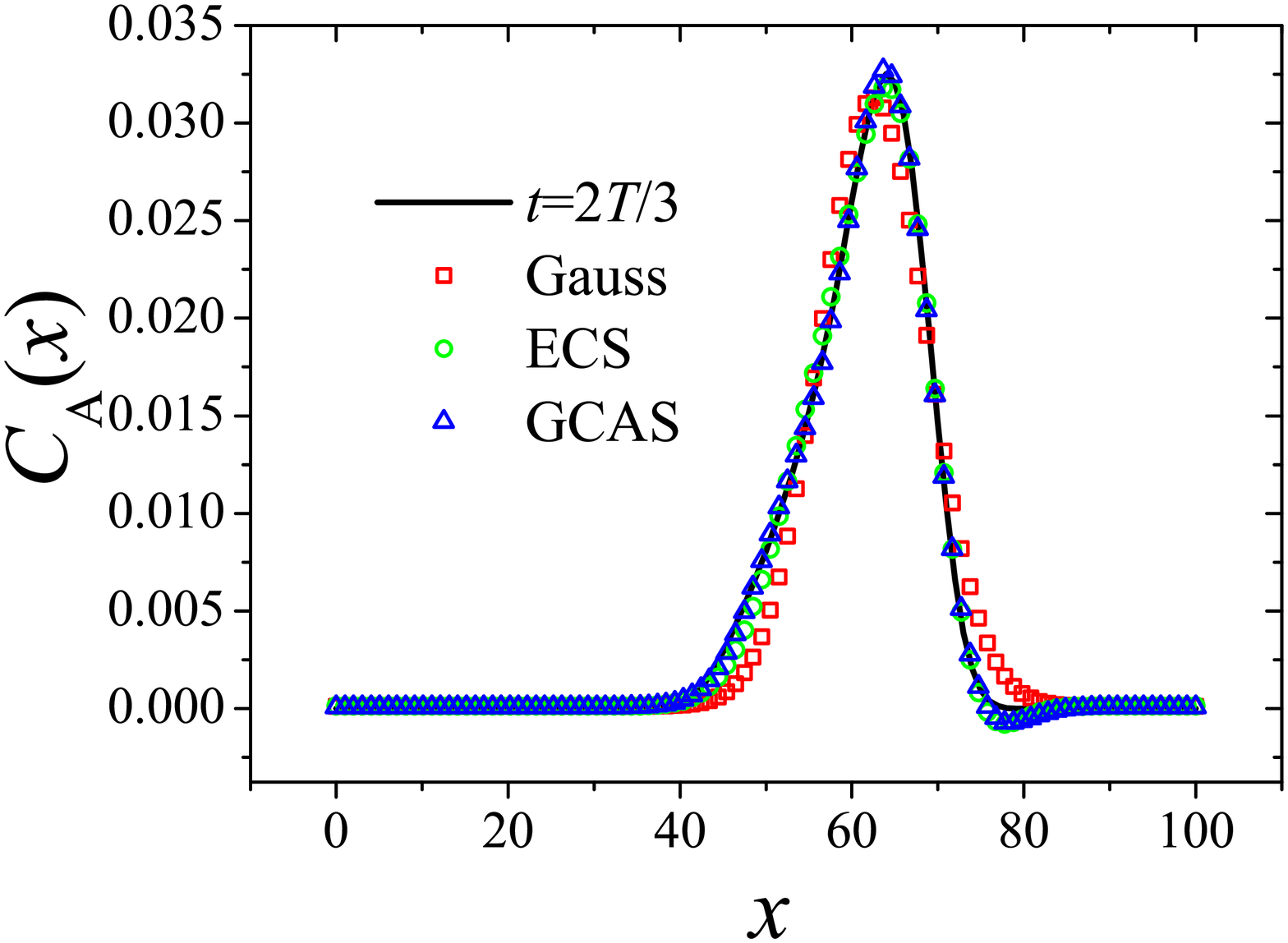} \includegraphics[width=0.75%
\columnwidth]{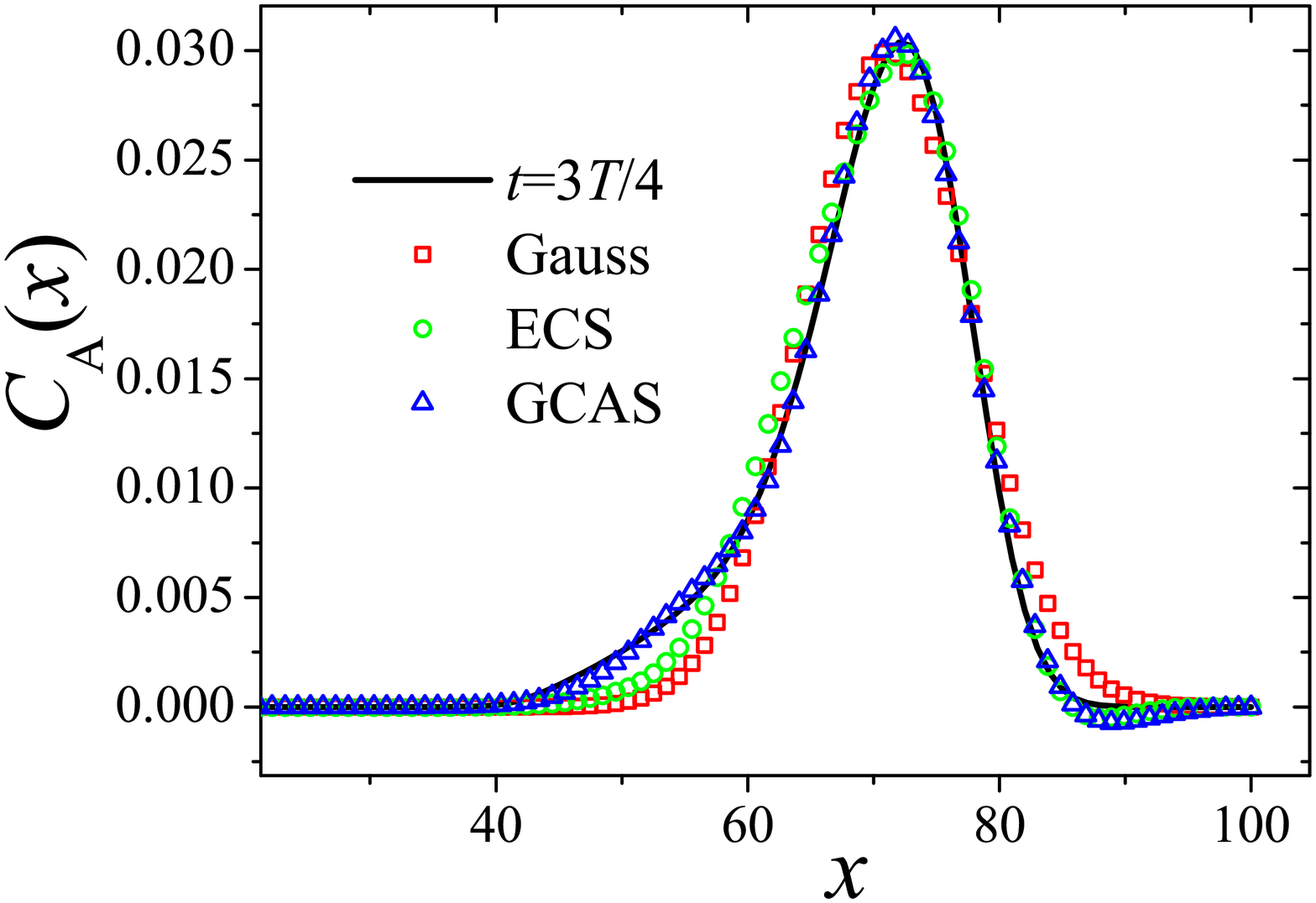}
\end{center}
\caption{Fits for particle density for the different times. We can see
deviation from Gaussian distribution chromatographic functions as ECS and
GCAS present better results although both use 3 parameters for fitting. }
\label{fitting}
\end{figure}

Finally we observe the possible fits for the particle distribution due to
nature of phenomena. So considering the larger resistance factor studied ($%
\alpha =0.4$) we fit the particle distribution at two different times: $%
t=2T/3$ and $t=3T/4$ where the effects of resistance can be better observed
(Fig. \ref{fitting}). We can check the deviation from normal behavior and
fits with two important chromatographic functions from literature were
checked:

1 - \textbf{Gram-Charlier peak function}

\begin{equation*}
f(x)=\frac{e^{-z^{2}/2}}{\sqrt{2\pi \sigma ^{2}}}\left[ 1+\frac{a_{1}}{3!}%
H_{1}(z)+\frac{a_{2}}{4!}H_{2}(z)\right]
\end{equation*}%
where $H_{1}(z)=z^{3}-3z$, $H_{2}(z)=z^{4}-6z^{3}+3$ with $z=(x-\mu )/\sigma 
$.

2 - \textbf{Edgeworth-Cramer peak function}%
\begin{equation*}
f(x)=\frac{e^{-z^{2}/2}}{\sqrt{2\pi \sigma ^{2}}}\left[ 1+\frac{a_{1}}{3!}%
H_{1}(z)+\frac{a_{2}}{4!}H_{2}(z)+\frac{10a_{1}^{2}}{6!}H_{3}(z)\right]
\end{equation*}%
where $H_{3}(z)=z^{6}-15z^{4}+45z^{2}-15$. In both cases, $\mu$ is the
center, $\sigma$ is the width and $a_{1}$ and $a_{2}$ are the unknown
parameters.

By using the Levenberg Maquardt method for non-linear regression we obtain
the convergent $R^{2}$ (i.e., the coefficient of determination) after the
iterations. We start by reporting the case where the interaction among
particles is less intense, i.e., $t=2T/3$. We obtain respectively for
Gaussian, ECS and GCAS: $0.9905$, $0.9987$ and $0.9977$. For $t=3T/4$ we can
observe a difference even bigger between the chromatographic ones and the
normal: $0.9703$, $0.9977$ and $0.9992$.

The deviation of gaussian transport occurs in many contexts including
chromotography \cite{Cromatograph} and noise flicker in semiconductors
devices \cite{noisesemiconductors, noisesemiconductors2}, for example. One
can suppose that a similar mechanism responsible for generating such
distributions could be related to our problem since that in transport
phenomena of molecules, or by thinking in electrons oriented by a field, a
similar resistance mechanism occurs: looking at the capture/emission of
electrons by traps given a Fermi level in semiconductor devices in the first
case, or by the capture of molecules and their reemission in the
chromatographic column as we reported in \cite{Cromatograph}.

So in this paper we propose a stochastic model to describe the movement of
particles against a contrary flux. This model is promising in order to
understand some peculiarities in the pedestrian dynamics alternatively to
the interesting social force models \cite{Helbing1995} since one analyzes
the non-normal behavior in particle density generated by interaction among
the particles. Such interaction was modeled in a simple way by a decreasing
term on the probabilities of a special random walk oriented by an `external
field' which does not allow the return of particles to previous cell. The
interesting phenomena of reduction/increasing of dispersion of the particles
is accentuated by the resistance term ($\alpha $) and a non-trivial behavior
for the symmetry and tail of distribution was monitored. We are sure that
such studies brings out an interesting class of stochastic process with
future universalities to be explored in the transport physics in random
mediums with applications in pedestrian dynamics under specific conditions.

\textbf{Acknowledgments --} This research was partially supported by the
Conselho Nacional de Desenvolvimento Cient\'{\i}fico e Tecnol\'ogico (CNPq),
under the grant 11862/2012-8 (R.S.). The work of A.A. was supported by Funda%
\c{c}\~{a}o de Amparo \`{a} Pesquisa do Estado de S\~ao Paulo (FAPESP),
under the grant 2013/22079-8.

\end{document}